\documentclass[10pt, journal]{IEEEtran}

\usepackage{cite}
\usepackage{amsmath,amssymb,amsfonts}
\usepackage{algorithmic}
\usepackage{graphicx,color}
\usepackage{subcaption}
\usepackage[table,xcdraw]{xcolor}
\usepackage{url}
\usepackage{hyperref}
\usepackage{textcomp}
\usepackage{flushend}
\usepackage[nobiblatex]{xurl}

\newcommand{\cb}{\color{black}}
\def\BibTeX{{\rm B\kern-.05em{\sc i\kern-.025em b}\kern-.08em
    T\kern-.1667em\lower.7ex\hbox{E}\kern-.125emX}}
\AtBeginDocument{\definecolor{ojcolor}{cmyk}{0.93,0.59,0.15,0.02}}
\def\OJlogo{\vspace{-4pt}\hskip-4pt\includegraphics[height=18pt]{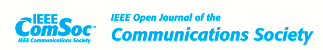}}

\begin{document}

\title{Highly Dynamic and Flexible Spatio-Temporal Spectrum Management with AI-Driven O-RAN: A Multi-Granularity Marketplace Framework}
 \author{\IEEEauthorblockN{Mehdi Rasti, \IEEEmembership{Senior Member,~IEEE}, Elaheh Ataeebojd,~\IEEEmembership{Member,~IEEE}, 
          Shiva Kazemi Taskooh,~\IEEEmembership{Member,~IEEE},
          \\ Mehdi Monemi,~\IEEEmembership{Member,~IEEE}, Siavash Razmi, and Matti Latva-aho, \IEEEmembership{Fellow,~IEEE}
		\thanks{M. Rasti, E. Ataeebojd, S. Kazemi Taskooh, M. Monemi, and  Matti Latva-aho are with the Centre for Wireless Communications, University of Oulu, Oulu, Finland (email: \{mehdi.rasti, elaheh.ataeebojd, shiva.kazemitaskooh, mehdi.monemi, matti.latva-aho\}@oulu.fi).
        }
        \thanks{
        S. Razmi is with the Faculty of Electrical and Computer Engineering, University of Tehran, Tehran, Iran (email: s.razmi@ut.ac.ir)}
		}
  }

  
\maketitle
\begin{abstract}
Current spectrum-sharing frameworks attempt to balance exclusive licensing and dynamic access but often fall short in adaptability. They are either static and long-term or insufficiently dynamic, operating on timescales of several minutes or longer, which restricts real-time adjustments. Moreover, these frameworks primarily focus on temporal spectrum sharing, overlooking the need for adaptive mechanisms across all temporal, spatial, and spectral dimensions.  
To address these limitations, we propose an adaptive, highly dynamic spectrum-sharing framework rooted in a marketplace approach. This framework integrates advanced artificial intelligence (AI) methods, including discriminative and generative AI (GenAI) techniques within the open-radio access network (O-RAN) architecture. It enables operators—including mobile, virtual, micro, and local operators—to forecast spectrum needs on a spatio-temporal basis, operating across multiple timescales (non-real-time, near-real-time, and real-time) and diverse 3D spatial granularities. Through this marketplace, managed by an authorized spectrum broker, operators can act as buyers or sellers based on their available and estimated spectrum needs.  
The O-RAN architecture supports both static (non-real-time) spectrum assignments and dynamic multi-timescale (near-real-time and real-time) spectrum trading markets. Advanced GenAI models enhance this framework by accurately predicting traffic variations, estimating spectrum demands, and allocating resources. This capability enables localized, time-specific 3D spatial spectrum leasing from brokers, optimizing resource utilization.  
The framework's modular, flexible, and open design surpasses existing spectrum management methods by fostering collaboration and innovation among operators, maximizing spectrum efficiency, and facilitating spectrum trading based on demand forecasts. This not only generates additional revenue but also reduces the costs associated with under-utilized spectrum.  
A critical research direction is determining allocation granularity and exploring the necessary spatio-temporal dynamics to advance beyond current static and insufficiently dynamic models.
\end{abstract}


\maketitle

\section{INTRODUCTION}
\IEEEPARstart{T}{he} rise of 6G and beyond communication systems, as outlined in ITU-R's 2030 framework \cite{ITU-20230}, has marked the beginning of a new era of connectivity, defined by immersive communication experiences, hyper-reliable low-latency services, and extensive support for massive device connectivity. These advancements are enabling a wide range of innovative applications, from enhanced mobile broadband (eMBB) for ultra-high data rates, to ultra-reliable low-latency communication (URLLC) for mission-critical applications, and massive machine-type communications (mMTC) to enable the Internet of Everything. Beyond traditional communication services, 6G is expected to support advanced use cases such as immersive extended reality, holographic communication, artificial intelligence (AI)-driven applications, and integrated sensing and communication for high-precision positioning, environmental monitoring, and intelligent transportation. These diverse use cases in vertical sectors, along with beyond connectivity services in 6G, require more spectrum to handle vast data exchanges and meet their varied connectivity needs.

However, efficiently managing the limited radio frequency spectrum remains a significant challenge in delivering these connectivity and beyond connectivity services and applications. Traditional static spectrum management methods are insufficient to meet the dynamic and diverse demands of modern communication networks because they operate on timescales of several minutes or longer, limiting real-time adjustments \cite{Ataeebojd2024}. Additionally, these frameworks often focus on spectrum sharing in only one or two of the spatial, temporal, and spectral dimensions, missing the potential efficiency and flexibility gains that spectrum sharing in a combination of all these dimensions could offer \cite{Net2022, ETSI_Spectrum}.

To address the challenges of current spectrum-sharing frameworks, an adaptive and highly dynamic marketplace-based spectrum-sharing framework managed by an authorized spectrum broker would be beneficial. This framework allows various operators, including mobile network operators (MNOs), mobile virtual network operators (MVNOs), micro-operators, and local operators \cite{Magazine2023}, to forecast their upcoming spectrum needs based on their internal demands across different spatial, temporal, and spectral dimensions and then participate in a multi-timescale marketplace (real-time [RT], near-real-time [near-RT], and non-real-time [non-RT]) as buyers or sellers, depending on their available and estimated spectrum need. This framework enables fine-grained, dynamic spectrum sharing, where operators capitalize on rapidly shifting statistical patterns of spectrum usage, which can change within milliseconds through flexible spectrum trading based on usage forecasts.

To deploy this adaptive and highly dynamic marketplace-based spectrum-sharing framework among various operators, it is essential to adopt a flexible, open, modular, and automated architecture with open interfaces, allowing seamless and real-time information exchange. The use of generative AI (GenAI) and open-radio access network (O-RAN) technologies enhances the capabilities of this framework. GenAI plays a pivotal role in our proposed framework by offering predictive analytics, traffic forecasting, spectrum estimation, and dynamic spectrum allocation \cite{GenAI-1}, allowing operators to lease spectrum from spectrum brokers in time-specific, localized 3D spatial, and spectral domains. GenAI also facilitates precise spectrum planning and rapid adaptation to fluctuating network demands and diverse quality of service (QoS) requirements \cite{GenAI-4}. On the other hand, O-RAN's open and softwarized architecture, along with its standardized interfaces that support multi-timescale decisions, exceed the capabilities of existing spectrum management methods, fostering collaboration and innovation within spectrum-sharing ecosystems \cite{Survey2023}. This architecture enables operators to engage in flexible and highly dynamic multi-timescale spectrum trade markets, with RT at sub-$10$ms intervals, near-RT for timescales between $10$ms and $1$s, and non-RT for timescales longer than $1$s \cite{dApps2022}. By integrating GenAI and O-RAN, the trading of spatio-temporally underutilized spectrum is optimized based on demand forecasts, generating additional revenue and reducing the costs associated with underutilized spectrum. 

\section{STATE OF THE ART AND MOTIVATION}
\subsection{SPECTRUM SHARING STATE-OF-THE-ART}
Traditional spectrum-sharing methods rely on long-term static licenses, granting exclusive access to specific services or operators. While this ensures interference-free access, it often leads to an underutilized spectrum. With increasing and fluctuating traffic demands in modern communication networks, these models struggle to adapt to real-time changes in spectrum availability, highlighting the need for more flexible and efficient solutions. Hence, cognitive radio (CR) \cite{CR2011} and dynamic spectrum access (DSA) \cite{DSA2007} technologies have emerged to address the growing demand for wireless resources in today's fast-evolving networks by enabling real-time adjustments and opportunistic use of underutilized spectrum.  In the CR model, temporary licensed users use advanced spectrum sensing to detect unused (white spaces) or underused (gray spaces) spectrum, allowing access when licensed users are inactive. DSA allows licensed users to lease portions of their spectrum to others on a short-term basis through ``spectrum-sharing agreements", facilitated by cooperative arrangements or auctions \cite{ETSI_2023, Tehrani2016}. 

With the advent of DSA and CR technologies, new spectrum-sharing frameworks have been introduced, such as licensed shared access (LSA), evolved licensed shared access (eLSA), and citizens broadband radio service (CBRS), which demonstrate the practical implementation of dynamic spectrum sharing in real-world scenarios. LSA allows additional licensed users to access spectrum already assigned to incumbents, administered by regulatory authorities. 
Unlike traditional spectrum leasing by MNOs, LSA ensures structured long-term sharing agreements for reliable and controlled spectrum access, offering an alternative to dynamic spectrum access (DSA). While LSA was successfully trialed in countries like Finland and France, particularly in the 2.3-2.4 GHz bands, it has not gained widespread global adoption. eLSA builds on LSA by enabling more dynamic and real-time spectrum adjustments, catering to complex sharing environments and high-connectivity networks, as explored in Italy. In the United States, CBRS uses a three-tier approach where incumbents, prioritized users, and general users share $150 $MHz of spectrum within the $3.55 $GHz to $3.7 $GHz range, with access coordinated by a spectrum access system (SAS) \cite{FCC, CBRS, FCC_spectrum}.

Additional techniques, such as dynamic frequency selection (DFS), listen before talk (LBT), and dynamic channel selection (DCS), are key components of efficient spectrum management strategies \cite{ETSI_Spectrum}. DFS dynamically selects frequency channels to avoid interference, particularly with radar systems, and is widely used in Wi-Fi. LBT, used in unlicensed spectrum bands, promotes fair spectrum access among multiple users by scanning the channel for ongoing transmissions before initiating new ones. DCS allows wireless devices to dynamically select the best available channel based on real-time conditions, minimizing interference and enhancing network performance. Table \ref{tab:comp} provides an overview of the existing key spectrum-sharing frameworks and techniques discussed in \cite{ETSI_Spectrum, ETSI_2023, CBRS, FCC_spectrum}. Furthermore, in \cite{IoT2024}, the authors proposed a spectrum-sharing mechanism between non-terrestrial and terrestrial networks within a marketplace framework, primarily focusing on facilitating cooperation between these network types through the O-RAN architecture. However, the implementation details of the spectrum marketplace approach in O-RAN were not considered in \cite{IoT2024}, nor was there an explicit focus on advanced AI techniques, such as GenAI.

\subsection{MOTIVATION}
Though approaches like LSA, eLSA, and CBRS provide a practical balance between exclusive licensing and dynamic, opportunistic access, they operate on timescales of several minutes or longer, which limits their flexibility and efficiency for real-time adjustments and opportunistic use of underutilized frequencies \cite{JSAC_2024}. Other frameworks, such as DFS, ITS, LBT, and DCS, primarily focus on temporal spectrum sharing and cannot address the growing need for adaptive sharing in other dimensions (such as 3D space)  that respond to millisecond-level changes in usage patterns. 
Therefore, this article proposes a more agile, automated, fine-grained, and highly dynamic spectrum marketplace, leveraging GenAI and the O-RAN technologies.

To address the limitations of current spectrum-sharing schemes, we need a highly flexible and dynamic spectrum marketplace, now made possible by AI-native O-RAN. By leveraging advanced AI and machine learning (ML) techniques and accelerators, operators can forecast traffic and associated spectrum on a spatio-temporal basis. Disaggregated spatio-temporal AI agents within O-RAN frameworks can then trade spectrum across various timescales, frequency, and 3D spatial locations, all supported by the softwarized, virtualized, and disaggregated architecture of O-RAN. This ensures that suitable frequencies can be automatically and swiftly selected, thereby maintaining continuous data communications and addressing the dynamic, time-sensitive spectrum requirements in deployed local 6G networks, even when elements of the local 6G network are located above ground and can move, as exemplified by the use of unmanned aerial vehicles (UAVs) whether as mobile base stations or user equipments, creating moving networks.
Moreover, GenAI and spectrum marketplace tasks can be executed by software applications known as xApps, rApps, and dApps, as well as control loops within the O-RAN architecture. This allows operators to participate as buyers or sellers in flexible and highly dynamic multi-timescale spectrum marketplaces based on their available and required spectrum. This approach facilitates seamless collaboration and innovation among operators, optimizing spectrum use in response to the diverse and fluctuating demands of modern networks.



\begin{table*}[!t]
\caption{Comparison of Existing Spectrum Sharing Frameworks and Techniques.}     \label{tab:comp}
\centering
\footnotesize
\setlength{\tabcolsep}{2pt}
\begin{tabular}{|>{\raggedright\arraybackslash}m{2.8cm}|>{\raggedright\arraybackslash}m{1.9cm}|>{\raggedright\arraybackslash}m{2.5cm}|>{\raggedright\arraybackslash}m{2cm}|>{\raggedright\arraybackslash}m{1.5cm}|>{\raggedright\arraybackslash}m{2.2cm}|>{\raggedright\arraybackslash}m{2.3cm}|>{\raggedright\arraybackslash}m{1.2cm}|}
\hline
\rowcolor{gray!25}
{\bf Framework/Technique} &{\bf Sharing dimension} & {\bf Deployment} & {\bf Spectrum allocation}  & {\bf Timescale} & {\bf Architecture} &{\bf Frequency Band} &{\bf Licensing Status}  \\ 
\hline\hline
LSA 
& frequency, geography &national \newline (in Europe) &manual,\newline before operation  & days  &centralized  & varies & licensed \\ \hline
\rowcolor{lightgray!15}
eLSA
& frequency, geography, time &local \newline (in Europe) & manual, \newline automated, during operation  & minutes & centralized  &varies & licensed \\ \hline
TVWS (Television White Spaces)&frequency, geography  &national \newline (in USA, UK, Kenya) &automated, \newline during operation  &hours &centralized  &TV White Spaces (Various) &unlicensed \\ \hline
\rowcolor{lightgray!15}
CBRS 
& frequency, geography  &national, local \newline (in USA)  &automated, \newline during operation  &minutes & centralized  (managed by SAS) &3.5 GHz (3.55 and 3.7 GHz) &licensed \\ \hline
Audio PMSE \newline(Programme Making and Special Events)&frequency,\newline geography,\newline time  &national, local \newline (in USA, UK, Australia) &manual,\newline before operation  &hours  &centralized  &470-694 MHz &licensed  \\ \hline
\rowcolor{lightgray!15}
NLL (National Local Licensing)&frequency,\newline geography  &national \newline (in USA, Europe) &manual,\newline before operation &hours  &decentralized  &varies &licensed  \\ \hline
DFS 
&time  &worldwide \newline (in USA, Europe) &contention-based  &milliseconds to seconds  &decentralized &5GHz (5.15-5.725 GHz) &unlicensed  \\ \hline
\rowcolor{lightgray!15}
LBT 
&time  &worldwide \newline (in USA, Europe) &contention-based  &milliseconds to seconds  &decentralized  &varies &unlicensed \\ \hline
Light Touch Leasing&frequency, geography  &local \newline (in USA, Europe) &automated,\newline during operation &minutes &centralized  &3.5 GHz (3550-3700 MHz) &licensed \\ \hline
\rowcolor{lightgray!15}
AFC \newline(Automated Frequency Coordination)&frequency, geography  & local \newline (in USA, Europe) &automated,\newline before operation  &days  &centralized  &6 GHz (5.925-7.125 GHz) &unlicensed \\ \hline
DCS 
&time  &worldwide \newline (in USA, Europe) &automated, \newline during operation  &milliseconds to seconds  &decentralized  &1.8 GHz (1710-1785 MHz) &licensed  \\ \hline
\rowcolor{lightgray!15}
AMBIT band  (Advanced Wireless Services for Broadband Innovation Technology)&frequency, geography  &national, local \newline (in USA) &automated, \newline during operation  &minutes  &centralized (managed by SAS)  &1695-2200 MHz  &licensed \\ \hline
AWS-3  (Advanced Wireless Services 3)&frequency, geography  &national, local \newline (in USA) &automated, \newline during operation  &minutes  &centralized  (managed by SAS) &1695-1710 MHz; 1755-1780 MHz; 2155-2180 MHz  &licensed \\ \hline
\rowcolor{lightgray!15}
\cite{IoT2024}  &frequency, geography, time  &national  &automated,\newline during operation  &milliseconds to seconds &centralized  &L-band,  Ka band \newline  &licensed \\ \hline
Our Proposed  Framework & frequency, geography, time & national, local & automated,\newline before operation,\newline during operation & milliseconds to seconds  &distributed/\newline centralized  &varies &licensed  \\ \hline
\end{tabular}
\end{table*}

\section{A FRAMEWORK FOR FLEXIBLE AND DYNAMIC MARKETPLACE-BASED SPECTRUM SHARING}
To achieve a flexible and dynamic marketplace-based spectrum-sharing system that addresses the aforementioned challenges, it should possess the following key characteristics:

\begin{itemize}
    \item[$\circ$] \textit{Multi-Granularity Management:} 
    The level of granularity in terms of time, space, and spectrum is crucial for enabling the flexibility needed to adjust spectrum management in response to dynamic network conditions, service lifecycles, and diverse QoS requirements. By incorporating multiple temporal, spatial, and spectral granularities, spectrum management can be structured to address the question of how granular and small a tuple of time-space-spectrum should be. This approach includes non-RT spectrum planning, as well as near-RT and RT spectrum adaptation, ensuring efficient resource allocation across 3D spaces and for a range of scenarios.
    
    \item[$\circ$] \textit{Highly Dynamic Spectrum Allocation:} 
    Spectral utilization varies significantly across five dimensions (5D)—frequency, time, and 3D space—based on factors such as the type of frequency band (sub-6GHz, centimeter-wave [cm-Wave], millimeter-wave [mm-Wave], sub-terahertz [THz], and potentially THz), as well as the spatial-temporal dynamics of network behavior, including environmental conditions, user mobility, and traffic demands. To address the challenges of adaptability and flexibility, spectrum allocation must be dynamically and automatically managed across different timescales. 
          
    \item[$\circ$] \textit{Flexible Spectrum Marketplace:} 
    With highly dynamic spectrum allocation, spectrum demand fluctuates significantly in spatial, temporal, and spectral dimensions, often compromising QoS when resources are insufficient. Operators require a spectrum marketplace to act as buyers to acquire additional spectrum during peak demand to meet QoS requirements and as sellers to lease excess spectrum during low-demand periods \cite{Magazine2023}. In multi-operator networks, spectrum marketplaces serve as dynamic, demand-driven platforms that enable spectrum trading among operators, optimizing both spectrum efficiency and QoS. This dynamic spectrum marketplace allows operators to participate as buyers or sellers depending on the tuple of time-space-spectrum, or even act as sellers in one tuple and buyers in another. This flexible spectrum marketplace can incentivize existing operators, encourage broader participation, and create new business opportunities in telecommunication, private networks, and Internet of things services \cite{Book2025}. To effectively manage dynamic spectrum allocation, the flexible marketplace must operate across multiple timescales—RT, near-RT, and non-RT modes.
\end{itemize}

\subsection{PROPOSED FRAMEWORK}
By incorporating these key requirements, we propose a flexible and dynamic spectrum-sharing marketplace that operates across multiple granularities and adapts to dynamic network conditions. Leveraging advanced AI and ML technologies for real-time decision-making, this marketplace is designed to address the complexities of modern communication networks, enhance spectral efficiency, and foster collaboration between new and established operators.

\begin{figure*}[!t]
\centering
\includegraphics[width=\textwidth]{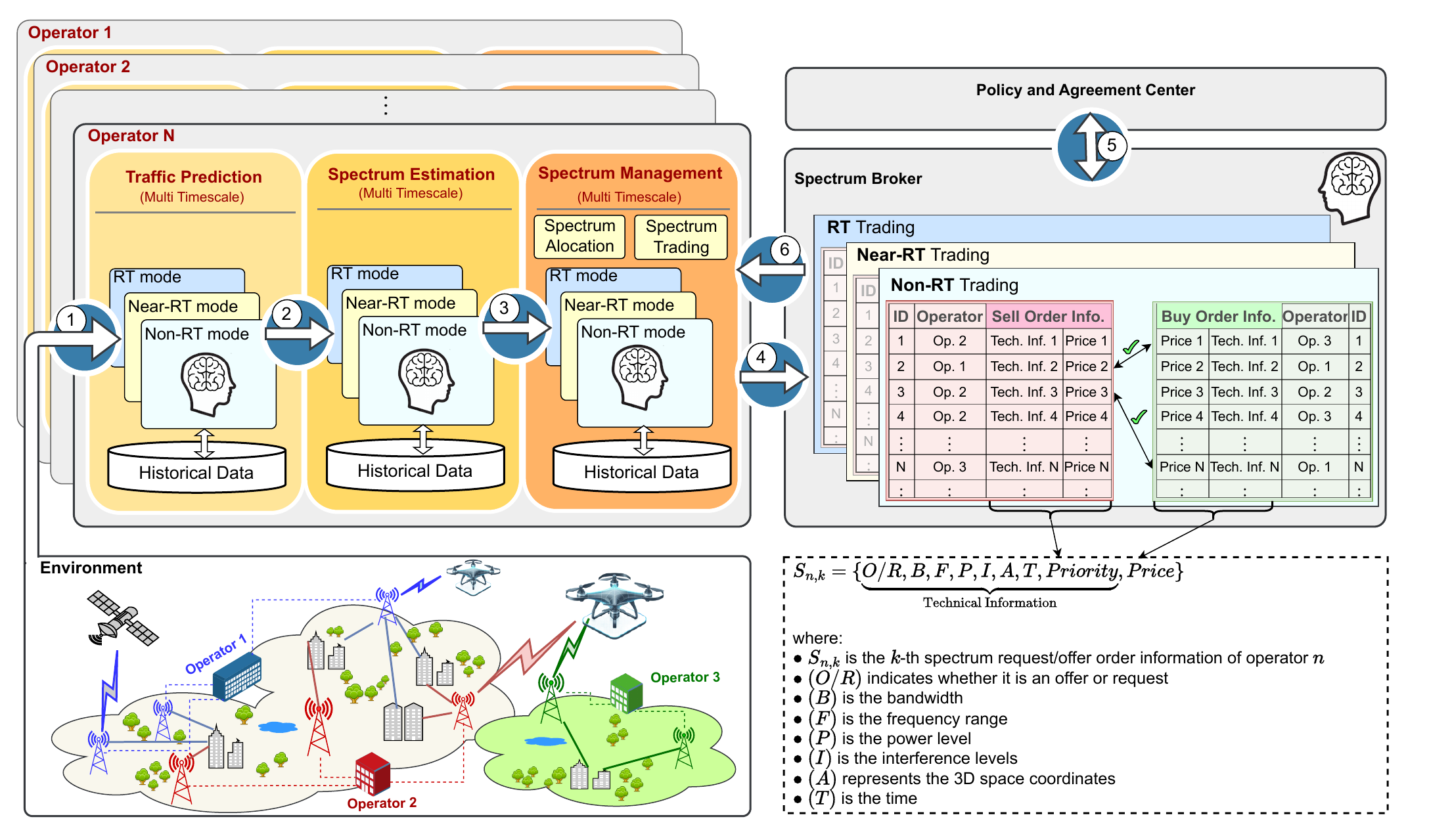}
\caption{A High-level Perspective of Our Proposed Dynamic and Flexible Marketplace-Based Spectrum-Sharing Framework. \label{fig:SSM}}
\end{figure*}

A high-level overview of our proposed marketplace-based spectrum-sharing framework is illustrated in Fig. \ref{fig:SSM}. In what follows, we will describe the various entities within this framework, their respective responsibilities, and their interactions.

\subsubsection{MAIN ENTITIES AND THEIR RESPONSIBILITIES}

 \begin{itemize}
     \item {\textit{Operator:}} Operators deliver a range of services to their subscribers and must effectively manage and allocate resources, particularly spectrum, to satisfy user demands. They act as either buyers or sellers, depending on whether they require additional spectrum or have excess to offer. Each operator needs to have the following entities:
     \begin{itemize}
     \item{\textit{Traffic Prediction:}} 
    This component captures spatial, temporal, and spectral dimensions of users' traffic demand, analyzing historical patterns and network traffic behavior to identify periods of high or low demand and specific spectrum requirements by leveraging AI and ML methods. To mitigate prediction errors arising from irregular traffic fluctuations, traffic prediction operates in a multi-timescale mode, providing the necessary flexibility to address unforeseen scenarios that were not previously anticipated. As a result, operators can directly manage their own spectrum or utilize the available spectrum based on their needs for upcoming time intervals.
    
    \item{\textit{Spectrum Estimation:}} 
    This entity evaluates the spectrum needed for anticipated data traffic while ensuring that QoS requirements are met. It considers key performance indicators (KPIs) such as peak data rate, latency, reliability, energy efficiency, spectrum efficiency, connectivity, and mobility. Additionally, it incorporates various key enabling techniques (KETs), including THz and mm-Wave communications, ultra-massive muli-input multi-output, reconfigurable intelligent surfaces, dynamic network slicing and virtualization, and non-orthogonal multiple access and its variants \cite{Rasti-6G}, all of which affect spectrum demand. Given the spatial-temporal-spectral dynamics of network behavior, employing spectrum estimation in a multi-timescale mode is crucial to minimize prediction errors and ensure the necessary flexibility.
   
    \item{\textit{Spectrum Management:}} 
    Operators evaluate their available spectrum to determine whether it meets the estimated spectrum requirements. If the available spectrum falls short, the operator, acting as a spectrum buyer, requests additional spectrum from the spectrum broker, providing details such as bandwidth, frequency range, power levels, interference levels, 3D spatial attributes, time, and price. Once the spectrum provisioning is complete, it is allocated to the subscribed users. Conversely, if the available spectrum exceeds the estimated requirements, the operator, in the role of a spectrum seller, temporarily offers the surplus spectrum and its associated characteristics to the spectrum broker.
    \end{itemize}
    
    \item {\textit{Spectrum Broker:}} 
    This component enables interactions between potential spectrum buyers and sellers across various timescales and manages the process of matching spectrum requests with available offers, with the aim of maximizing the number of successful matches.   To facilitate this process, the broker can operate either through a centralized entity (i.e., a third party) or via decentralized ledger technologies such as blockchain \cite{blockchain-1}.  

    \item \textit{Policy and Agreement Center:} 
    This center periodically gathers the latest information from the spectrum broker, directs decisions regarding changes in spectrum demand for the multi-timescale market model, specifies the actions to be taken by the spectrum broker, and incorporates trading agreements. 
 \end{itemize}

\subsubsection{OPERATIONAL WORKFLOW}
The process of defining spectrum requests and offers, as well as trading them, involves the following steps, as shown in Fig. \ref{fig:SSM}. 

\textit{Step 1:} Operators continuously monitor traffic conditions to analyze traffic flow, service types, the number of users, and external factors such as the time of day or special events across various timescales.

\textit{Step 2:} Utilizing AI and ML algorithms, real-time and historical data traffic are analyzed for pattern recognition and anomaly detection. To capture irregular traffic fluctuations arising from the spatial-temporal-spectral dynamics of the network, traffic predictions are performed across multiple temporal, spatial, and spectral granularities in a multi-timescale approach. This facilitates accurate forecasts, effective management of network fluctuations, and preparation for uncertainties.

\textit{Step 3:} Taking into account KPIs and KETs, future spectrum demand is estimated across various temporal, spatial, and spectral granularities in a multi-timescale mode. This enables operators to self-adapt their spectrum requirements based on traffic and usage patterns.

\textit{Step 4:} After assessing the available spectrum against the estimated requirements, operators engage in a flexible spectrum market across different timescales, acting as either spectrum buyers or sellers. Buyers request additional spectrum, while sellers offer surplus spectrum. Given the fluctuating spectrum demand across five dimensions, operators can simultaneously submit multiple offers and requests, including the necessary technical information, as depicted in Fig. \ref{fig:SSM}.

\textit{Step 5:} The spectrum broker collaborates with the policy and agreement center to enforce the necessary policies and to match offers and requests based on the provided information.

\textit{Step 6:} Following the spectrum broker's decisions, both the matched spectrum seller and buyer are notified. Additionally, details about the matches are forwarded to the policy and agreement center for record-keeping.

Our proposed flexible, marketplace-based spectrum-sharing framework is a dynamic, demand-responsive platform that enables spectrum trading among operators, maximizing resource efficiency across the network \cite{Book2025}. This optimization is achieved through advanced ML and AI-driven techniques for real-time traffic prediction and spectrum management, designed to handle unexpected fluctuations. With multi-granularity adjustments, the framework can adapt spectrum allocation in response to changing network conditions, service lifecycles, and diverse QoS needs. However, the current RAN infrastructure lacks the automation and virtualization needed for the seamless onboarding of new operators and flexible, on-demand resource allocation. O-RAN, with its open, agile, and disaggregated architecture, supports efficient information exchange within flexible spectrum marketplaces. Utilizing open interfaces and RAN intelligent controllers (RICs), O-RAN enables our framework to operate across multiple timescales, creating an automated, data-driven, zero-touch platform that simplifies operator onboarding and resource management across RAN environments. 

Integrating GenAI with O-RAN further enhances these capabilities, adding intelligence and adaptability for various RAN functions, including data analysis, traffic forecasting, anomaly detection, model training, spectrum demand prediction, and dynamic spectrum sharing. This combination creates a more resilient, efficient, and responsive spectrum marketplace that meets the demands of diverse services and adapts smoothly to fluctuating traffic loads.
In the following section, we provide a detailed explanation of the GenAI-empowered O-RAN for our proposed spectrum-sharing framework. 

\begin{figure*}[!t]
\centering
\includegraphics[width=\textwidth, height=13 cm]{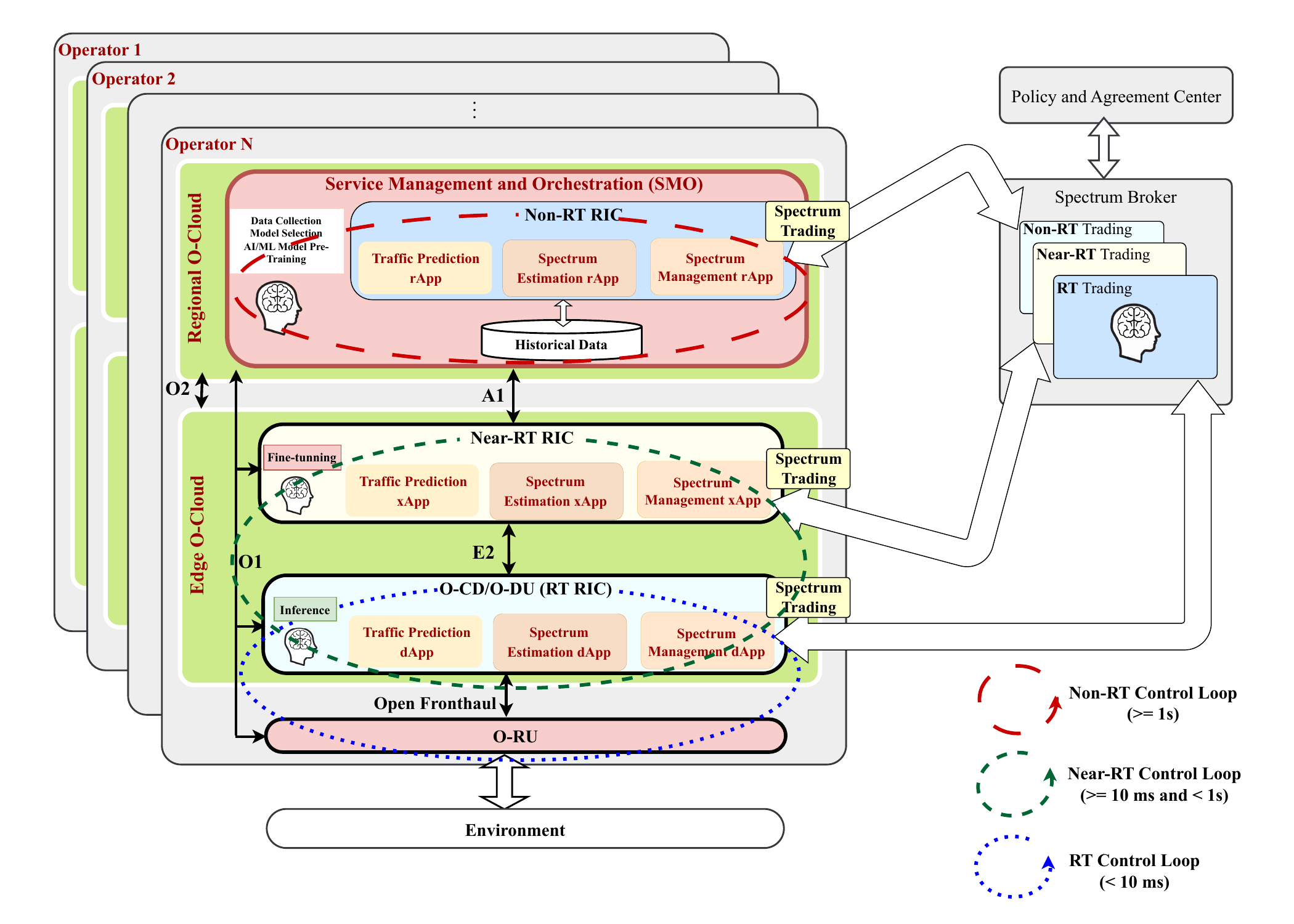}
\caption{GenAI-Empowered O-RAN for Spectrum Sharing through Marketplaces. \label{fig:framework}}
\end{figure*}

\section{GENAI-EMPOWERED O-RAN FOR MARKETPLACE-BASED SPECTRUM SHARING}
The GenAI-empowered O-RAN architecture for marketplace-based spectrum sharing is illustrated in Fig. \ref{fig:framework}. In what follows, we delve deeper into the specifics of O-RAN and GenAI technologies, highlighting their roles and synergies within this framework. 

\subsection{EMPOWERING THE PROPOSED FRAMEWORK WITH O-RAN} 
O-RAN is a highly adaptable, modular, open, and intelligent architecture designed to enhance interoperability and flexibility within RAN. It achieves this by leveraging standardized interfaces (such as E2, A1, and O1) and incorporating AI-driven automation. The primary components and benefits of O-RAN include \cite{Survey2023}:
\begin{itemize}
    \item  
   Non-RT RIC: operates at the service management and orchestration (SMO) level, addressing tasks for long-term spectrum sharing on timescales longer than one second. It collects and analyzes network data to train AI models for predicting spectrum demand, facilitates spectrum sharing among operators, develops long-term policies for spectrum allocation, monitors network conditions and adapts policies, and generates insights to help operators assess and adjust spectrum needs.
    \item 
   Near-RT RIC: operates closer to the radio network, managing tasks that demand rapid responses on timescales between $10 $ms and $1$s. It empowers dynamic spectrum allocation based on real-time network demands, enables rapid spectrum sharing for short-term needs, adapts policies to optimize spectrum usage, and enhances user experience through seamless connectivity, optimized handovers, and improved reliability via interference management, load balancing, and mobility management.    
    \item {O-CU, O-DU, and O-RU:} open-centralized unit (O-CU) manages higher-layer functions within the RAN protocol stack and handles QoS, acting as a centralized controller to optimize resource utilization in spectrum sharing. The open-distributed unit (O-DU) coordinates real-time processing and scheduling, adjusting resources dynamically to meet varying demands for efficient shared spectrum use. Meanwhile, the open-radio unit (O-RU) is responsible for low-level physical layer and radio frequency functions.
    \item rApps, xApps, and dApps: rApps function within the non-RT RIC, focusing on long-term spectrum-sharing optimization through AI-driven traffic forecasting and policy-based resource management. In contrast, xApps are utilized in the near-RT RIC, managing short-term spectrum-sharing activities such as monitoring near-RT demand and adjusting spectrum usage among operators to alleviate network congestion and enhance spectral efficiency. They facilitate rapid customization and optimization of spectrum resources to address immediate, location-specific requirements. dApps are integrated into O-CUs and O-DUs, utilizing decentralized resources to improve spectrum sharing at sub-$10$ms timescales \cite{dApps2022}. They promote collaborative decision-making among operators, enabling them to share real-time insights and data regarding spectrum usage and conditions.
    \item O-Cloud: is a cloud computing platform consisting of a pool of physical infrastructure nodes tailored to meet O-RAN requirements and support virtualized network functions. These nodes can be distributed across various locations, such as cell sites, edge clouds, and regional clouds. The O-Cloud facilitates the deployment and management of RAN components, allowing for scalable and flexible network operations.
    \item SMO: is responsible for the end-to-end management and orchestration of network services, with a particular emphasis on spectrum sharing among operators. It coordinates the lifecycle management of network functions and services, utilizing data analytics and AI/ML to optimize and automate spectrum management. By implementing zero-touch automation, the SMO minimizes manual intervention, enhances operational efficiency, and dynamically adapts to real-time network conditions, thereby supporting the flexible and responsive characteristics of O-RAN.
\end{itemize}

\subsection{EMPOWERING THE PROPOSED FRAMEWORK WITH  GENERATIVE AI} 
GenAI is a specialized area of AI focused on creating new content, such as text, images, and video, by learning patterns and distributions from existing data. This unique capability makes GenAI a valuable tool for dynamic spectrum sharing in wireless networks \cite{5G-Americans}.
The application of GenAI in wireless networks can be broadly divided into two key capabilities: generation and representation. These capabilities can be leveraged individually or in combination with discriminative AI to address challenges like data traffic forecasting, spectrum estimation across various time scales, and spectrum management. The details are as follows:
\begin{itemize}
    \item[$\circ$] \textit{Generation Capabilities}: GenAI can generate synthetic data to serve as input for discriminative AI models. This includes creating synthetic data to augment training datasets, particularly when real-world data is scarce or costly to obtain. GenAI can simulate diverse traffic patterns and user behaviors to improve data traffic forecasting. It can also generate various spectrum usage scenarios under different conditions, such as peak and off-peak hours or across geographical regions, to support spectrum estimation. This ability to generate large volumes of synthetic data is crucial for simulations and for providing data to discriminative AI models.
    \item[$\circ$] \textit{Representation Capabilities}: GenAI can be utilized for data compression and feature selection, leading to data dimensional reduction, which enhances data storage efficiency and transmission while boosting the performance of discriminative AI models.  For instance, it can simplify complex datasets of traffic and spectrum usage into more manageable and compact formats.  These compressed datasets can then be used by discriminative AI models for tasks such as classification, prediction, and decision-making, significantly improving efficiency and accuracy. Moreover, GenAI facilitates the selection of the most relevant features for model training, further optimizing the effectiveness and performance of discriminative AI models for spectrum sharing in wireless networks. 
\end{itemize}

\subsubsection{THE POTENTIALS OF GENAI IN THE SPECTRUM MARKETPLACE} 
\ \\
The capabilities of GenAI help address challenges like data traffic forecasting, spectrum estimation, and spectrum management in wireless networks \cite{5G-Americans}. We can benefit from GenAI capabilities for spectrum sharing across the following domains:
\begin{itemize}
    \item[$\circ$] \textit{Traffic Forecasting}: GenAI can play a transformative role in traffic forecasting and spectrum allocation by leveraging its unique capabilities to predict demand spikes and dynamically optimize resources.  Specifically, by generation capability, GenAI can simulate diverse, previously unseen scenarios, such as a 6G network sharing spectrum with incumbent services, to predict optimal times and locations for spectrum utilization. By employing models like generative adversarial networks (GANs) and variational autoencoders (VAEs), GenAI can generate synthetic traffic scenarios that reflect varying demand patterns, user behaviors, and network conditions \cite{GenAI-Traffic}. Additionally, its representation capabilities allow for compressing and encoding complex historical data into compact representations, facilitating efficient analysis and improved model performance. These compressed representations can be utilized to train discriminative AI models, enhancing their accuracy in forecasting traffic loads and optimizing network resources.  For instance, VAEs can reduce high-dimensional data into latent spaces that preserve essential features, enabling better prediction of congestion hotspots and more efficient spectrum allocation strategies.
    \item[$\circ$] \textit{Spectrum Management}: GenAI offers innovative solutions to mitigate spectrum scarcity by generating optimal sharing agreements and dynamic allocation schemes.  In scenarios involving multiple operators sharing spectrum, such as urban areas with limited resources, GenAI can create dynamic allocation models that adjust spectrum access based on real-time network load and demand predictions. This ensures equitable use and optimal performance for all stakeholders.  While GenAI facilitates fair and efficient spectrum sharing, challenges such as interference management, equipment compatibility, fair sharing policies, QoS, licensing, and revenue sharing in shared infrastructures require additional considerations. GenAI can provide foundational insights to address these complexities effectively.
    \item[$\circ$] \textit{Dynamic Spectrum Sharing}: GenAI facilitates flexible and adaptive spectrum allocation through its ability to dynamically generate insights and models based on real-time demand and usage patterns. Unlike discriminative AI, GenAI can synthesize spectrum usage scenarios, predict future demand, and create adaptive allocation strategies tailored to current network conditions.
    \item[$\circ$] \textit{Spectrum Estimation}: GenAI revolutionizes spectrum estimation by leveraging its generative capabilities to predict spectrum usage and availability with high precision. By analyzing massive historical and real-time data, GenAI can estimate the spectrum needs of various applications and services, ensuring optimal allocation while minimizing interference. This capability is particularly advantageous in dynamic environments like urban areas with dense network deployments, where spectrum demand can fluctuate rapidly.  For example, generative diffusion models (GDMs) can create high-fidelity traffic data that captures intricate patterns and temporal variations. This synthetic data is invaluable for modeling traffic in complex environments, such as urban networks with dense user populations, where accurate spectrum predictions are critical for effective spectrum management and resource optimization \cite{spectrumestimation}.
\end{itemize}

\subsubsection{VARIOUS PHASES OF GENAI IN ORAN}
\ \\ 
Applying GenAI models for spectrum sharing involves several key phases: data collection, pre-training, fine-tuning, and inference. Each phase plays a crucial role in ensuring the models are effective and adaptable to varying network conditions.
\begin{itemize}
    \item[$\circ$] \textit{Data Collection}: This phase involves gathering historical data to train GenAI models. Given the limited availability of relevant data, GenAI can augment datasets by generating synthetic data.  Moreover, pre-processing is essential to maintain data quality and diversity, while dimensionality reduction can improve model accuracy and optimize storage. Besides, data comes from diverse sources, such as users, operators, and base stations, and spans multiple modalities like spectrum usage, traffic patterns, and users' mobility. Leveraging GenAI models that can handle multimodal data enhances performance. Collected data includes information on users' demand, traffic patterns, and spectrum usage across various time scales, such as RT, near-RT, and non-RT. 
    \item[$\circ$] \textit{Pre-training}: In this phase, GenAI models are trained on massive datasets to learn foundational patterns and features. Given that pre-training requires significant computational resources and time, it necessitates implementation in the rApps module of the non-RT RIC in the O-RAN architecture.
    \item[$\circ$] \textit{Fine-tuning}: While pre-trained models provide a strong foundation, they often require fine-tuning for specific tasks and time scales, such as spectrum trading.  Fine-tuning involves adapting the model with data specific to RT and near-RT time scales or particular tasks, ensuring improved accuracy and task relevance.  Besides, continuous data collection and updates are critical to keep the models responsive to environmental and network changes.  For RT and near-RT applications, fine-tuning is best performed in the xApps module of the near-RT RIC in O-RAN, enhancing adaptability to dynamic network conditions.
    \item[$\circ$] \textit{Inference}: The final phase involves using fine-tuned GenAI models to make decisions at different time scales.  To achieve optimal performance for RT and near-RT tasks, deploying these models in the dApps module of O-RAN is recommended. This ensures rapid and efficient decision-making aligned with real-time network demands. 
\end{itemize}
\subsection{THREE CONTROL LOOPS FOR SPECTRUM MARKETPLACE IN GENAI-EMPOWERED O-RAN}
Our proposed spectrum-sharing framework incorporates three control loops within O-RAN, facilitating zero-touch spectrum management and sharing, as shown in Fig. \ref{fig:framework}. In what follows, we provide a detailed explanation of each control loop.
\begin{itemize}
    \item \textit{Non-RT Closed-Loop:} The non-RT closed loop addresses strategic, long-term spectrum-sharing functions. Through the O1/A1 interfaces, the non-RT RIC gathers historical spectrum demand data for users across regions, using this information to train AI/ML models that capture spatial, temporal, and spectral patterns in user traffic. This allows for analysis of historical trends and network traffic behavior. Considering KPIs and KETs, future spectrum demand is predicted over broader temporal, spatial, and spectral scales within the non-RT timeframe, enabling operators to plan, manage policies, and allocate resources based on traffic and usage trends. Ultimately, the non-RT closed loop supports operators in updating spectrum distribution to O-CUs based on long-term demand insights.
    \item \textit{Near-RT Closed-Loop:}  In this loop, the near-RT RIC gathers and processes real-time data from the network via the O1/E2 interfaces, fine-tunes the pre-trained AI/ML models by the non-RT RIC and applies them to identify patterns, predict network conditions, and optimize resource allocation.  Targeting key functions such as interference management, load balancing, and mobility management, the near-RT closed loop enables proactive adjustments to maintain network performance and QoS, with spectrum provisioning at a finer granularity. This approach helps the network respond quickly to fluctuations in user demand and environmental changes, ensuring stability and efficiency. The seamless integration of near-RT and non-RT closed loops provides comprehensive network management, balancing strategic goals with immediate operational demands.
    \item \textit{RT Closed-Loop:} The RT closed loop is dedicated to ultra-fast decision-making, managing the most time-sensitive network tasks to maintain critical functions with maximum responsiveness. It uses high-speed, low-latency interfaces to continuously monitor network conditions, user behavior, and environmental factors, allowing for instantaneous data processing to detect anomalies and forecast immediate network demands. Enabled by dApps for real-time inference and control, the RT closed loop can react instantly to changes, reducing latency, minimizing packet loss, and ensuring smooth user experiences. In seamless coordination with the near-RT and non-RT loops, it elevates overall network responsiveness and adaptability, supporting efficient and high-quality service delivery.
\end{itemize}

These closed loops collaborate to create a highly responsive and adaptive spectrum-sharing framework. This framework adjusts to varying conditions, ensuring efficient spectrum utilization and optimized performance to meet the dynamic demands of modern communication technologies. It also maintains QoS across all scenarios.

\section{FUTURE RESEARCH DIRECTIONS}
The proposed framework presents several promising research directions that are crucial for the advancement of dynamic spectrum sharing. These pathways are essential not only for the theoretical development of the model but also for its practical implementation and optimization in real-world settings. While not exhaustive, these research directions highlight key areas that must be addressed to unlock the framework's full potential. Each direction offers unique challenges and opportunities, necessitating a multidisciplinary approach that integrates expertise from communication theory, economics, machine learning, and network optimization.

\subsubsection{ROBUST TRAFFIC PREDICTION AND SPECTRUM ESTIMATION FOR COMMUNICATION AND EMERGING SERVICES}  
A key research direction involves developing robust methods to forecast intermittent data traffic for services like eMBB, mMTC, URLLC, and emerging applications, such as AI and sensing. This entails creating temporal-spatial-spectral traffic demand predictions that accurately estimate demand requirements for upcoming time slots. The challenge lies in accommodating diverse service requirements and deployment scenarios, such as differences in indoor versus outdoor traffic patterns. Moreover, translating forecasted traffic into spectrum needs is another important research challenge.
Advanced ML, AI, and GenAI methods provide valuable predictive capabilities for synthesizing future traffic scenarios, enabling robust traffic forecasting even under highly dynamic conditions and fluctuations.

\subsubsection{OPTIMAL GRANULARITY IN SPECTRUM ALLOCATION}
Determining the ideal granularity across spatial, temporal, and spectral dimensions is a critical research direction, leveraging AI to balance detailed allocation with operational efficiency. The objective is to achieve an optimal level of granularity that meets service requirements effectively, ensuring precise spectrum management without adding unnecessary complexity to the process.

\subsubsection{BLOCKCHAIN}
Currently, spectrum trading is often facilitated through centralized regulatory authorities or intermediaries, such as brokers \cite{blockchain-1}. However, this approach presents several challenges, including risks of a single point of failure, trust issues, and security vulnerabilities. Decentralized ledger technologies, such as blockchain, offer a potential solution by securely recording spectrum trading transactions, including spectrum allocations and payments, in a transparent and tamper-resistant manner \cite{blockchain-2}.  
To enhance ownership traceability, scalability, robustness, and trust within the proposed spectrum-sharing framework, integrating blockchain and smart contract technologies can help secure, automate, and streamline the process of defining and executing spectrum trades among operators. By reducing reliance on intermediaries such as brokers, banks, and regulators, this approach can lower transaction costs while ensuring clear service-level agreements for spectrum trading. However, for effective implementation, the consensus mechanism must be carefully designed to accommodate the varying timescales and operational requirements of the framework.

\subsubsection{FEDERATED LEARNING}
Federated learning enables multiple operators to collaboratively train shared AI models using decentralized data, enhancing spectrum management without the need to exchange sensitive user information. This method not only improves privacy and security but also facilitates the development of more accurate and robust AI models that are better suited to varying network conditions.

\subsubsection{DYNAMIC PRICING AND INCENTIVIZATION}
Research in this domain will investigate economic and behavioral strategies to provide price signals and incentives to operators, effectively managing peak demand while promoting shifts in temporal, spatial, and spectral traffic. By encouraging users to move their traffic away from peak periods, this approach aims to balance the overall network load, enhance efficiency, and sustain service quality. Achieving this requires a deep understanding of customer behavior and the development of dynamic pricing models that motivate operators to adjust their usage and encourage end users to alter their traffic patterns, ultimately resulting in a more balanced and optimized spectrum utilization.

\subsubsection{GLOBAL SPECTRUM SHARING REGULATIONS}
As spectrum marketplaces enhance the buying and selling of spectrum resources, international regulatory oversight becomes essential to ensure compliance, prevent disruptions, foster cooperation, promote transparency, and support cross-border operations. To effectively address these challenges, a robust regulatory framework is required to facilitate efficient and fair spectrum sharing on a global scale, encourage innovation, and guarantee reliable communication services.

\begin{figure}[!t]
    \centering
    \includegraphics[width=1\columnwidth, height=6cm]{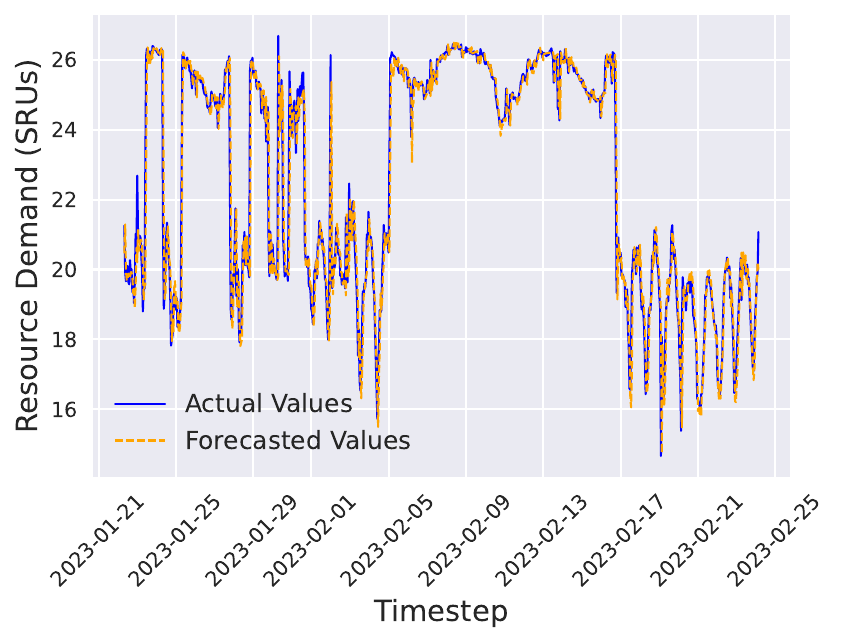}
    \caption{Actual and forecasted resource demand.}
    \label{fig:forecast}
\end{figure}

\begin{figure}[!t]
    \centering
    \begin{subfigure}[b]{1\columnwidth}
        \centering
        \includegraphics[width=\columnwidth, height=6cm]{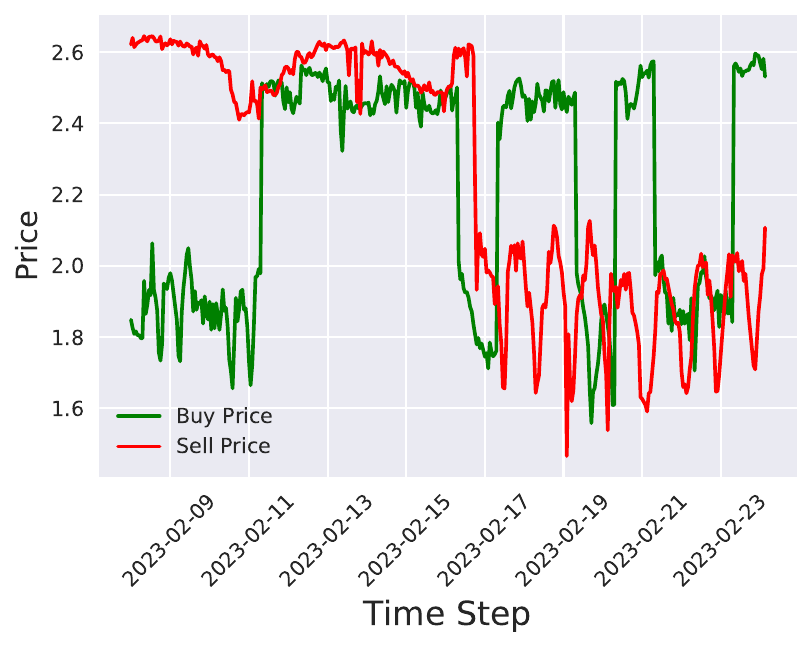}
        \caption{Granularity = 1 hour}
        \label{fig:price1}
    \end{subfigure}
    \vfill
    \begin{subfigure}[b]{1\columnwidth}
        \centering
        \includegraphics[width=\columnwidth, height=6cm]{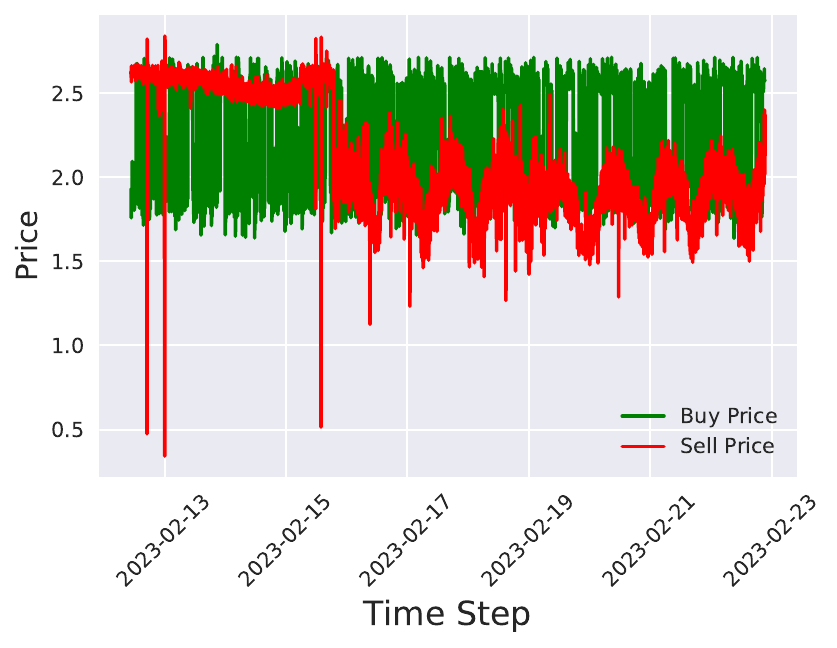}
        \caption{Granularity = 1 minute}
        \label{fig:price2}
    \end{subfigure}
    \caption{Price of selling and buying SRUs corresponding to (a) one-hour and (b) one-minute time granularities.}
    \label{fig:price}
\end{figure}

\section{NUMERICAL RESULTS}

In this section, we present numerical results to illustrate the financial benefits that network operators can gain from the proposed AI-driven spectrum market. Operating within the O-RAN framework and leveraging (generative) AI-driven procedures, this marketplace enables operators to place bid orders to buy or ask orders to sell spectrum resources efficiently across different time scales. While here we focus on non-RT spectrum marketplace transactions facilitated through rApps, similar results can also be extended to near-RT or RT levels.

We consider a network setup with two operators: one providing LTE services and the other offering new radio (NR) services. We consider the real-world LTE traffic data collected through BladeRF software-defined radio (SDR) and online watcher of LTE (OWL) where the hourly averaged traffic data is sourced from \cite{gopal2024adapshare} for the timespan of 2023-01-22 to 2023-02-22. For the NR operator, the traffic data used to calculate the market price was generated using the TimeGAN generative model, trained on the LTE operator data. The first 15 days of data were used for model training, while the remaining days were reserved for evaluation. 
Our focus here is on the financial revenue of the LTE network (referred to as the target network), obtained through the buying or selling of resources. The networks share a pool of physical resource blocks called here as spectrum resource units  (SRUs), with network demand quantified as the average number of SRUs required. To forecast the LTE network's traffic demands, we employed Amazon's Chronos model \cite{ansari2024chronos}, a pretrained generative large language model (LLM) for time series forecasting. Using a zero-shot approach, we input the previous 72 hours of traffic data to predict the next hour's demand. 
Fig. \ref{fig:forecast} compares the predicted and actual traffic demands. 
Based on these forecasts, the target operator generates bid or sell orders in the spectrum marketplace. A predefined threshold of 24 SRUs is considered here. If demand exceeds this threshold, a bandwidth deficit arises, leading to packet drops and user dissatisfaction. Conversely, when demand falls below the threshold, unused resources are wasted. To address this, a decision-making model determines whether to buy, sell, or take no action after receiving the forecast for the next hour.

The agent implements a double deep Q-network (DDQN) architecture, utilizing two identical neural networks; an online network for action selection and a target network for action evaluation. Each network consists of three fully connected layers: an input layer accepting 4-dimensional state vectors (current bandwidth, forecast of next time stamp $t+1$, moving average of past $n$ forecast error, moving average of past $n$ demand values), followed by two hidden layers of 128 neurons each with ReLU activation functions, and an output layer producing Q-values for 5 possible actions. The agent maintains an experience replay buffer of size 100,000 to store state transitions and sample batches of 64 experiences for training, optimizing the networks using the Adam optimizer with a learning rate of 0.001. Stability is enhanced through soft target network updates with $\tau=0.005$. The agent employs an $\epsilon$-greedy exploration strategy where $\epsilon$ decays from 1.0 to 0.01 at a rate of 0.995, balancing exploration and exploitation during training.
At each timestep, the agent selects the action $a$ from a set of allowed actions $\mathcal{A}$ including buying or selling 1.5 or 3 SRUs or taking no action. This translates in \(a \in \mathcal{A}\), where \(\mathcal{A} = \{-3, -1.5, 0, 1.5, 3\}\) SRUs. Negative actions correspond to ask orders (selling), while positive actions correspond to bid orders (buying). For simplicity, it is assumed that all orders are executed instantaneously. 
The SRU sell and buy prices are based on a linear coefficient of traffic demand in the NR 5G and LTE operator. We have considered that price is obtained as $P=c \times T$, where $P$ is the price, $c$ is a constant considered as 0.1, and $T$ is the traffic demand. 
Considering the network traffic, the price of selling and buying SRUs at different granularities is depicted in Fig. \ref{fig:price}.

\begin{figure}[!t]
    \centering
    \includegraphics[width=\columnwidth, height=6cm]{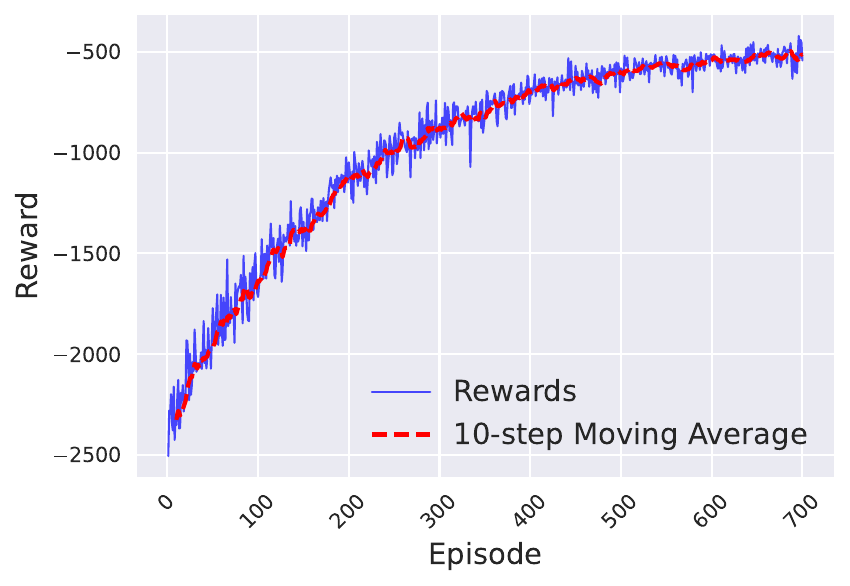}
    \caption{Reward values over training episodes.}
    \label{fig:reward_per_episode}
\end{figure}

\begin{figure}[!t]
    \centering
    \begin{subfigure}[b]{1\columnwidth}
        \centering
        \includegraphics[width=\columnwidth, height=6cm]{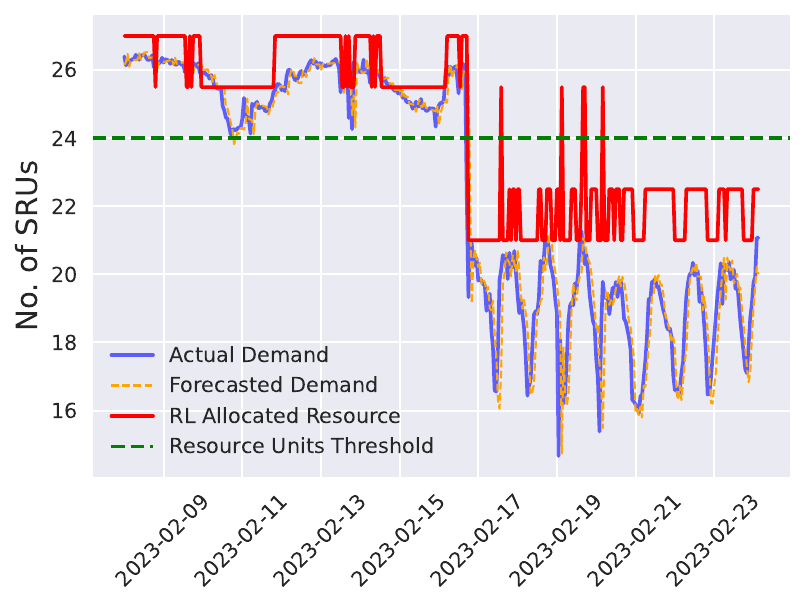}
        \caption{Granularity = 1 hour}
        \label{fig:resource1}
    \end{subfigure}
    \vfill
    \begin{subfigure}[b]{1\columnwidth}
        \centering
        \includegraphics[width=\columnwidth, height=6cm]{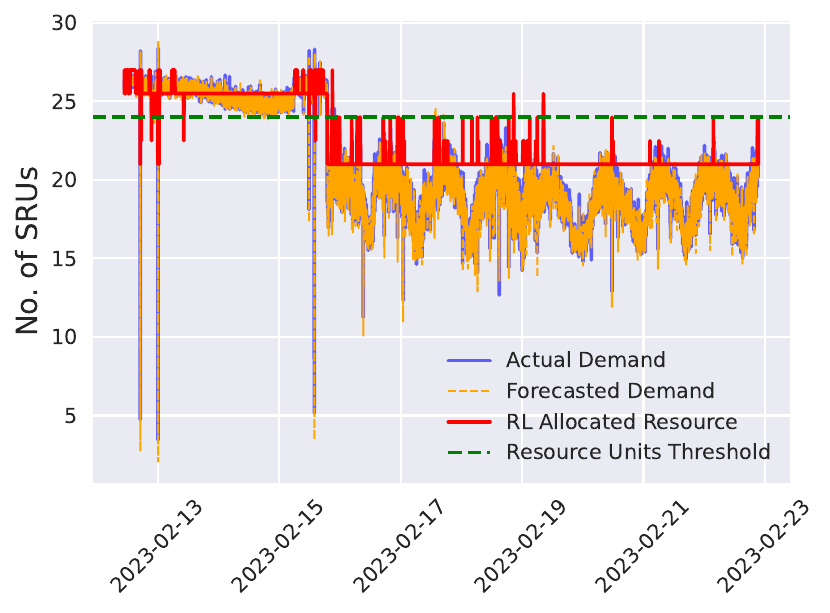}
        \caption{Granularity = 1 minute}
        \label{fig:resource2}
    \end{subfigure}
    \caption{Resource demand/allocation in the validation phase for time granularity of (a) one hour and (b) one minute.}
    \label{fig:resource_price_per_time}
\end{figure}

The decision-making model is guided by a reward function 
characterized as follows:
\begin{align*}
    R = - (\alpha D + \beta S + \gamma P + T),
\end{align*}
where \(D\) is the resource deficit (per SRU), \(S\) is the resource surplus (per SRU), \(P\) is the monetary cost (positive for buying, negative for selling), and \(T\) is the transaction cost.  We have set the parameters coefficients as \(\alpha = 8\), \(\beta = 2\) and \(\gamma = 0.5\), prioritizing user satisfaction over profit. 
Fig. \ref{fig:reward_per_episode} shows the reward values over training episodes. 

\begin{figure}[!t]
    \centering
    \begin{subfigure}[b]{1\columnwidth}
        \centering
        \includegraphics[width=\columnwidth, height=6cm]{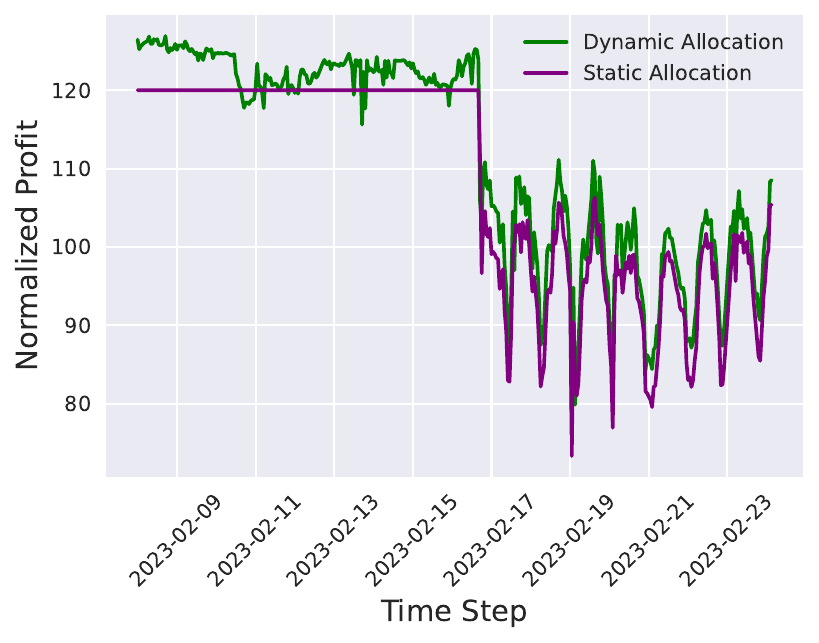}
        \caption{Granularity = 1 hour}
        \label{fig:profit1}
    \end{subfigure}
    \vfill
    \begin{subfigure}[b]{1\columnwidth}
        \centering
        \includegraphics[width=\columnwidth, height=6cm]{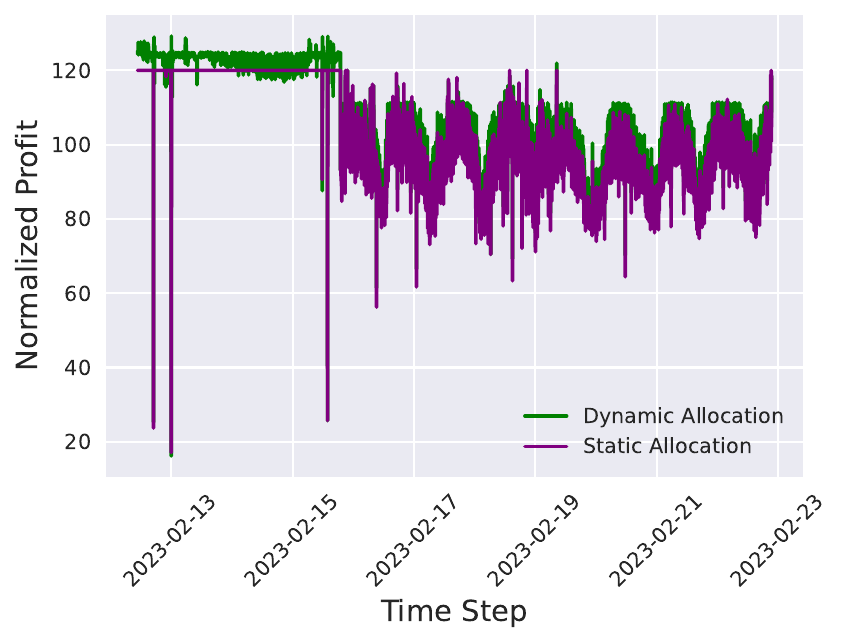}
        \caption{Granularity = 1 minute}
        \label{fig:profit2}
    \end{subfigure}
    \caption{Financial profit of the target operator in the validation phase for time granularity of (a) one hour and (b) one minute.}
    \label{fig:profit}
\end{figure}

After training the model, its performance was evaluated on the validation set, both at on-hour and one-minute granularities. Fig. \ref{fig:resource_price_per_time} highlights the resource demand and allocation trends, while Fig. \ref{fig:profit} depicts the normalized financial profit achieved by the target operator in the proposed marketplace compared to the conventional static scenario. Initially, as seen in Fig. \ref{fig:resource_price_per_time}, the network demand consistently exceeded the available resources. The model responded by purchasing additional resources to meet user requirements. In contrast, a static model with a fixed SRU threshold would drop excess packets, resulting in user dissatisfaction and missed opportunities to benefit from the extra demand. From 2023-02-17 onward, the network demand fell below the threshold. At this stage, the model began selling surplus bandwidth to generate profit, whereas the static model would let unused resources go to waste. 
The proposed dynamic resource allocation model consistently demonstrated higher financial profits, as illustrated in Fig. \ref{fig:profit}, highlighting its capability to adapt to fluctuations in network demand. The superior performance during the surplus phase stems from the profit generated by selling unused resources. Similarly, the enhanced performance during the deficit phase is attributed to the higher revenue gained by serving additional users, which outweighs the cost incurred for purchasing extra resources from the other operator.

\section{CONCLUSION}\label{Con}

This article has outlined a marketplace-based spectrum-sharing framework designed to enable operators to meet diverse and variable spectrum demands, maximizing resource utilization and responding to requirements within milliseconds. By incorporating GenAI and O-RAN technologies, this framework supports a more flexible, automated, and fine-grained approach to spectrum sharing across spatial, temporal, and spectral dimensions. Future research could enhance this framework by integrating additional advanced technologies and investigating new use cases, driving further improvements in spectral efficiency and network performance. As we transition toward a highly connected and dynamic 6G landscape, the advancement of adaptive and intelligent spectrum management frameworks will be essential for meeting the ultra-low latency, high reliability, and massive connectivity requirements of next-generation communication systems.

\end{document}